# Electronic, magnetic properties and correlation effects in the layered quaternary iron oxyselenide $Na_2Fe_2Se_2O$ from first principles

D. V. Suetin, I. R. Shein,* and A. L. Ivanovskii

*Institute of Solid State Chemistry, Ural Branch of the Russian Academy of Sciences,*

*Ekaterinburg, GSP-145, 620990, Russia*

**Abstract**

By means of the first-principle calculations, we have investigated electronic, magnetic properties and correlation effects for the newly discovered layered oxyselenide $Na_2Fe_2Se_2O$. Our results reveal that the electron correlations in the Fe 3*d* bands promote a transition of $Na_2Fe_2Se_2O$ from magnetic metallic or half-metallic states to the antiferromagnetic Mott-insulating state. In addition, the bonding picture in $Na_2Fe_2Se_2O$ is described as an anisotropic mixture of ionic and covalent contributions.

Quasi-two-dimensional (2D) Fe-based oxides, especially after the discovery of high-$T_C$ superconductivity in electron-doped LaFeAsO, [1] have received tremendous attention as new promising superconducting materials. Besides, these layered multi-component systems with unique structural and chemical flexibility exhibit a rich variety of fascinating physical properties such as spin-density-wave behavior, various types of magnetic order *etc*., see reviews. [2-8]

The effects of electron correlations on the properties of these materials are the subjects of intensive current discussion. [5,6] So, for the iron oxychalcogenides $La_2O_2Fe_2S_2O$ and $La_2O_2Fe_2Se_2O$ with an expanded Fe-square lattice, there is evidence [9] that the correlation effects promote the Mott-insulating state.

Very recently, the new tetragonal (space group *I4/mmm*, # 139) oxyselenide $Na_2Fe_2Se_2O$ was successfully prepared. [10] This phase includes the same building blocks [$Fe_2Se_2O$] as the aforementioned $La_2O_2Fe_2Se_2O$, Fig. 1; however, for $Na_2Fe_2Se_2O$ these blocks are separated along the $c$ axis by double layers of ions $Na^+$ (blocks [$Na_2$]) – instead of [$La_2O_2$] blocks for $La_2O_2Fe_2Se_2O$. This leads to a considerable decrease in the spacing of blocks [$Fe_2Se_2O$]/[$Fe_2Se_2O$] (the parameter $c$ = 14.641 Å for $Na_2Fe_2Se_2O$ [10] *versus* $c$ = 18.605 Å for $La_2O_2Fe_2Se_2O$ [9]); at the same time, the square lattice of iron atoms for $Na_2Fe_2Se_2O$ is even more expanded ($a$ = 4.107 Å [10]) than for $La_2O_2Fe_2Se_2O$ ($a$ = 4.085 Å [9]). The available experiments [10] reveal that $Na_2Fe_2Se_2O$ undergoes an antiferromagnetic (AFM) transition below $T_N$ ~ 73K and adopts a gap with the energy of 0.26 eV.

In this Brief Report, we focus on the interplay of electronic, magnetic properties and correlation effects in the newly discovered iron oxyselenide $Na_2Fe_2Se_2O$ and show that the electron correlations in the Fe 3*d* states are responsible for the formation of the antiferromagnetic Mott-insulating state of this material.

Since no data about the electronic structure of $Na_2Fe_2Se_2O$ are available, as far as we know, it seems helpful to begin the discussion with the general features of the electronic spectrum and inter-atomic binding for this system in the paramagnetic state. This analysis was performed by means of the full-potential linearized augmented plane wave method with mixed basis APW+lo (FLAPW) implemented in the WIEN2k suite of programs. [11] The generalized gradient approximation (GGA) to exchange-correlation potential in the PBE form [12] was used. The basis set inside each muffin tin (MT) sphere was split into core and valence subsets. The core states were treated within the spherical part of the potential only and were assumed to have a spherically symmetric charge density in MT spheres. The valence part was treated with the potential expanded into spherical harmonics to $l = 4$. The valence wave functions inside the spheres were expanded to $l = 12$. We used the corresponding atomic radii: 2.00 a.u. for Na, 1.90 a.u. for Fe, 2.10 a.u. for Se, and 1.70 a.u. for O atoms. The plane-wave expansion was taken to $R_{MT} \times K_{MAX}$ equal to 7, and the *k* sampling with 10×10×10 *k*-points in the Brillouin zone was used. The calculations were considered to be converged when the difference in the total energy of the crystal did not exceed 0.1 mRy as calculated at consecutive steps.

The results for $Na_2Fe_2Se_2O$ as calculated without spin polarization are presented in Fig. 1 and Table I. We find that the Fe 3*d* electrons contribute mainly to the DOS near the Fermi energy $E_F$, whereas the O 2*p* states are centered at ~ 6.0 eV below the $E_F$. In turn, the Se 4*p* states lie between the O 2*p* and Fe 3*d* states. We see the overlapping of the Fe 3*d* - Se 4*p* and Fe 3*d* - O 2*p* states, which form directional Fe-Se and Fe-O bonds inside blocks [$Fe_2Se_2O$]. These covalent bonds are well visible in Fig. 2, where the valence charge density maps in characteristic sections of $Na_2Fe_2Se_2O$ are depicted.

In turn, the adjacent blocks [$Fe_2Se_2O$] and Na sheets are coupled mainly by ionic bonds, though weak overlapping of Na-O valence orbitals is visible in Fig. 2. The preliminary picture of ionic bonding comes from a simple ionic model, where the assignment of the usual oxidation numbers of atoms gives immediately the ionic formula $Na^{1+}{}_2Fe^{2+}{}_2Se^{2-}{}_2O^{2-}$. This implies that the adjacent blocks adopt the same charges $[Na_2]^{2+}/[Fe_2Se_2O]^{2-}$, *i.e.* the charge transfer (2*e*) occurs from blocks $[Na_2]^{2+}$ to blocks $[Fe_2Se_2O]^{2-}$. To estimate the actual atomic charges and the inter-block charge transfer numerically, the Bader's scheme [13] was applied. This yields the effective atomic charges: $Na^{0.814+}{}_2Fe^{0.766+}{}_2Se^{1.003-}{}_2O^{1.149-}$ and the inter-block charge transfer ~ 0.81*e* per f.u. Thus, we conclude that the inter-atomic bonding in $Na_2Fe_2Se_2O$ can be described as an anisotropic mixture of covalent and ionic contributions, where inside blocks [$Fe_2Se_2O$] covalent Fe-Se and Fe-O bonds take place (besides, inside these blocks the ionic interactions occur owing to charge transfer Fe → Se and Fe → O), whereas between blocks [$Na_2$]/[$Fe_2Se_2O$] mainly the ionic bonds emerge.

Coming back to the electronic spectrum of $Na_2Fe_2Se_2O$, let us emphasize that $E_F$ falls within a sharp peak of Fe 3*d* states, *i.e.* the non-magnetic solution leads to a metallic-like state, contradicting with the experiment. [10]

Therefore, to establish the role of spin coupling in the formation of the insulating state of this material, we extended our calculations to various magnetically ordered structures of $Na_2Fe_2Se_2O$. Note also that clear evidence of itinerant magnetism for this material comes directly from NM calculations within the Stoner's criterion: $N^{3d}(E_F)\cdot I > 1$, where $N^{3d}(E_F)$ is the density of 3*d* states at the Fermi level (in states/eV·atom·spin) and *I* is a so-called internal exchange parameter; for 3*d* metals the *I* value is about 0.7.[14] As a result, the critical $N^{3d}(E_F)$ value is ~ 1.4 states/eV·atom·spin. Thus, the calculated high value of $N^{Fe\ 3d}(E_F)$, see Table II, shows that $Na_2Fe_2Se_2O$ should be magnetic.

Then, to explore the interplay between the magnetic structure and the insulating behavior of $Na_2Fe_2Se_2O$, at first we have examined the simplest ferromagnetic spin ordering.

In Fig. 3, we plot the total and projected density of states for FM $Na_2Fe_2Se_2O$. As expected, spin polarization occurs primarily within the Fe 3*d* bands. As a result, in the spin-down channel, $E_F$ falls within partially filled Fe 3*d*↓ bands with a nonzero density of carriers at the Fermi level, whereas for the reverse spin channel a gap at about 1.3 eV appears. Thus, the FM solution yields the *magnetic half-metallic* behavior of $Na_2Fe_2Se_2O$, for which the spin density polarization at the Fermi level is $P = \{N\downarrow(E_F) - N\uparrow(E_F)\}/\{N\downarrow(E_F) + N\uparrow(E_F)\} = 1$ and the total magnetic moment (per cell) adopts an integer value: $MM^{tot} = 8.0\ \mu_B$. As to the local

MMs, besides the maximal magnetic moments on the Fe atoms ($MM^{Fe}$ ~ 3.3 $\mu_B$ per atom), additional quite small *induced* MMs associated with Se ($MM^{Se}$ ~ 0.18 $\mu_B$ per atom) and oxygen ($MM^{O}$ ~ 0.28 $\mu_B$ per atom) are formed owing to strong hybridization of Fe 3*d* - Se 4*p* and Fe 3*d* - O 2*p* orbitals, respectively.

Our next step is related to the antiferromagnetic solution. Though the available experiments [10] exhibit an AFM behavior of $Na_2Fe_2Se_2O$ with $T_N$ ~ 75K, no exact long-range spin ordering is known. Therefore, in our calculations we searched for the ground state spin ordering among three different collinear commensurate in-plane AFM spin configurations (AFM1-AFM3), which belong to the most typical forms of spin alignment in iron-based layered crystals, reviews.[5-8] Here, for AFM1 (so-called stripe-like configuration), the magnetic structure within each [$Fe_2Se_2O$] block consists of chains of antiparallel Fe spins that are coupled ferromagnetically together, with antiferromagnetic ordering between adjacent [$Fe_2Se_2O$]/[$Fe_2Se_2O$] blocks. For AFM2 (so-called checkerboard configuration), the magnetic structure within each [$Fe_2Se_2O$] block consists of chains of parallel Fe spins that are coupled antiferromagnetically; in turn, spin ordering is also antiferromagnetic between adjacent [$Fe_2Se_2O$]/[$Fe_2Se_2O$] blocks. Finally, for AFM3, within each [$Fe_2Se_2O$] block the Fe spins are coupled ferromagnetically, with antiferromagnetic ordering between adjacent [$Fe_2Se_2O$]/[$Fe_2Se_2O$] blocks.

The results indicate (Table II) that the AFM3 ordering has the lowest energy, whereas the other antiferromagnetic configurations, AFM1 and AFM2, have energies higher than the FM state. For AFM3, the local magnetic moments on the Fe atoms are ~ 3.29 $\mu_B$ per atom. But for all of the AFM configurations, $Na_2Fe_2Se_2O$ also remains metallic-like, with nonvanishing spin density at the Fermi level, Table II.

Thus, all the previous results point to enhanced correlation effects in the formation of the gap for $Na_2Fe_2Se_2O$.

To check this issue, we applied the GGA+$U$ technique [15] for two most stable magnetic configurations (AFM3 and FM) of $Na_2Fe_2Se_2O$. Our results reveal that in this case the FM state of $Na_2Fe_2Se_2O$ retains a half-metallic behavior, whereas the required gap appears for AFM3. Besides, for the on-site Coulomb repulsion parameter $U = 5$ eV and Hund's parameter $J = 1$ eV, the value of the band gap (~ 0.30 eV) agrees well with the experimental estimations: ~ 0.26 eV. [10]

Thus, we see that $Na_2Fe_2Se_2O$ demonstrates a Mott-insulating behavior and belongs to a rather rare group [9,16,17] of quasi-two-dimensional correlation-induced insulators with antiferromagnetic spin ordering. But, unlike the related systems $La_2O_2Fe_2S_2O$ and $La_2O_2Fe_2Se_2O$ with the checkerboard-like AFM configuration, [9] for $Na_2Fe_2Se_2O$ the most favorable AFM ordered state was found when the Fe spins within each [$Fe_2Se_2O$] block are coupled ferromagnetically together with antiferromagnetic ordering between adjacent [$Fe_2Se_2O$]/[$Fe_2Se_2O$] blocks. At the same time, the energies of the competing configurations (stripe-like and checkerboard-like) differ by less than 0.02 to 0.06 eV. Certainly, the neutron experiments seem very desirable to clarify up this issue.

In summary, we studied the newly synthesized iron oxyselenide $Na_2Fe_2Se_2O$ theoretically to understand the electronic, magnetic properties and correlation effects in this quasi-two-dimensional system. We found that the Fe 3$d$ electronic bands determine the main peculiarities of the aforementioned properties, providing the formation of AFM ordered state and the correlation effects, which promote the Mott-insulating state of this material.

---

*Corresponding author; e-mail: shein@ihim.uran.ru

TABLE I. The optimized lattice parameters (*a*, *b,* and *c*, in Å) and cell volumes (V, in $Å^3$) for non-magnetic (NM), ferromagnetic (FM), and three antiferromagnetic (AFM) phases of $Na_2Fe_2Se_2O$.

| phases * | *a* | *b* | *c* | V |
|---|---|---|---|---|
| NM | 3.9280 | - | 14.6956 | 113.37 |
| | (4.107) | | (14.641) | (123.48) |
| FM | 4.0929 | - | 14.8088 | 124.04 |
| AFM1 | 4.0923 | 4.0859 | 14.8188 | 123.89 |
| AFM2 | 4.0505 | 4.0403 | 14.5170 | 118.79 |
| AFM3 | 4.1025 | - | 14.8937 | 125.33 |

* experimental results [10] are given in parentheses; for the types of AFM order (AFM1-3) see the text.

TABLE II. Total and Fe 3*d* densities of states at the Fermi level ($N(E_F)$, in states/eV/f.u.), atomic magnetic moments (μ, in $\mu_B$), and total-energy differences (ΔE, in eV/atom) for non-magnetic (NM), ferromagnetic (FM), and three antiferromagnetic (AFM) phases of $Na_2Fe_2Se_2O$.

| phases * | $N^{tot}(E_F)$ | $N^{Fe\ 3d}(E_F)$ | $\mu^{Fe}$ | $\mu^{Se}$ | $\mu^{O}$ | ΔE |
|---|---|---|---|---|---|---|
| NM | 12.663 | 9.708 | - | - | - | 0 |
| FM | $0^{\uparrow}/2.383^{\downarrow}$ | $0^{\uparrow}/1.801^{\downarrow}$ | 3.299 | 0.176 | 0.282 | -0.284 |
| AFM1 | $2.309^{\uparrow}/2.537^{\downarrow}$ | $1.784^{\uparrow}/1.959^{\downarrow}$ | 3.108 | 0 | 0 | -0.266 |
| AFM2 | $1.267^{\uparrow}/1.256^{\downarrow}$ | $0.921^{\uparrow}/0.922^{\downarrow}$ | 3.091 | 0 | 0 | -0.227 |
| AFM3 | $1.347^{\uparrow}/1.346^{\downarrow}$ | $1.011^{\uparrow}/1.010^{\downarrow}$ | 3.289 | 0.164 | 0.276 | -0.288 |

* for the types of AFM order (AFM1-3) see the text.

## FIGURES

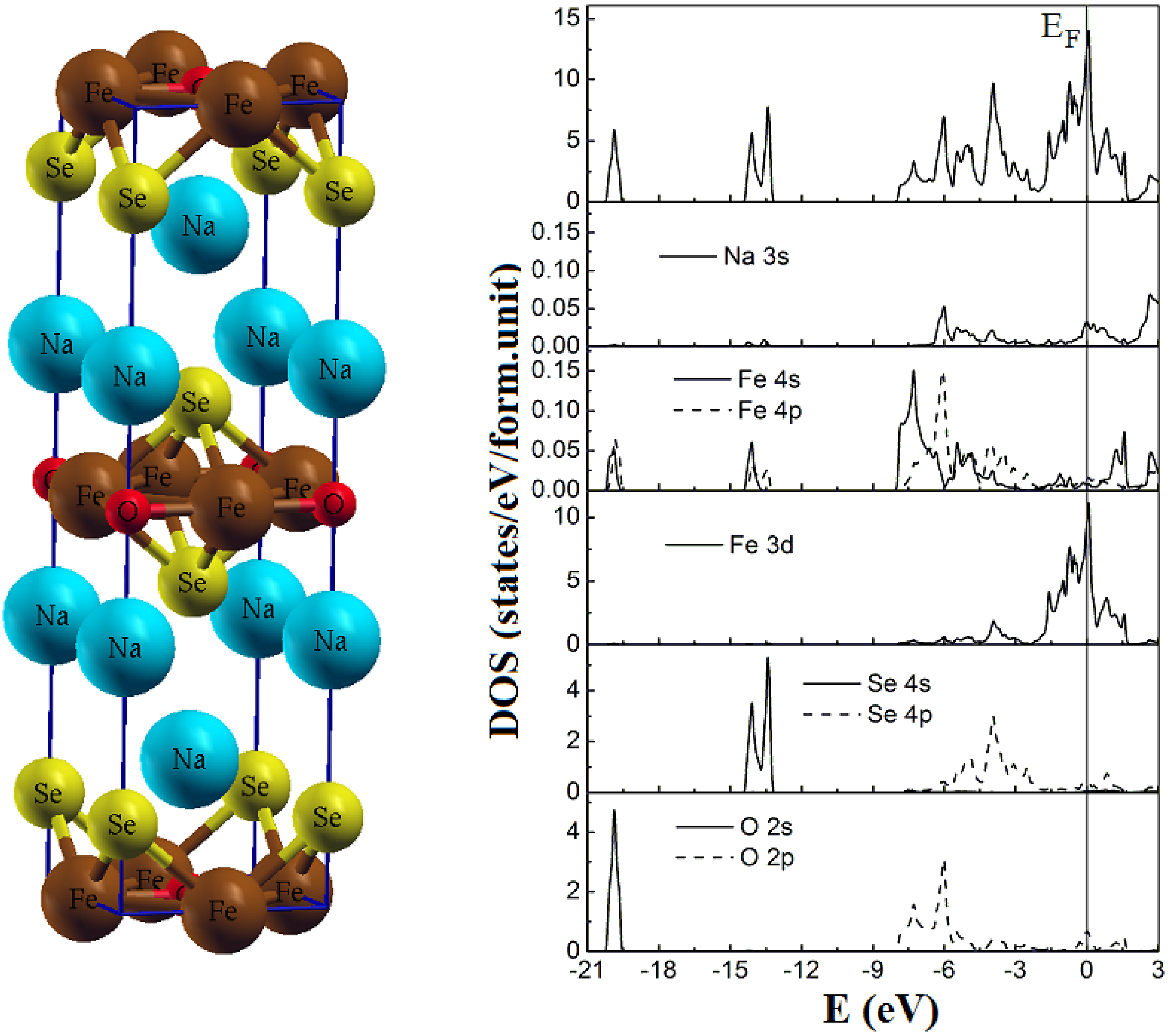

FIG. 1. (*Color online*). (*Left panel*) Crystal structure of $Na_2Fe_2Se_2O$; (*right panel*) densities of states of the non-magnetic phase of $Na_2Fe_2Se_2O$ as obtained within GGA. The Fermi level is given at 0.0 eV.

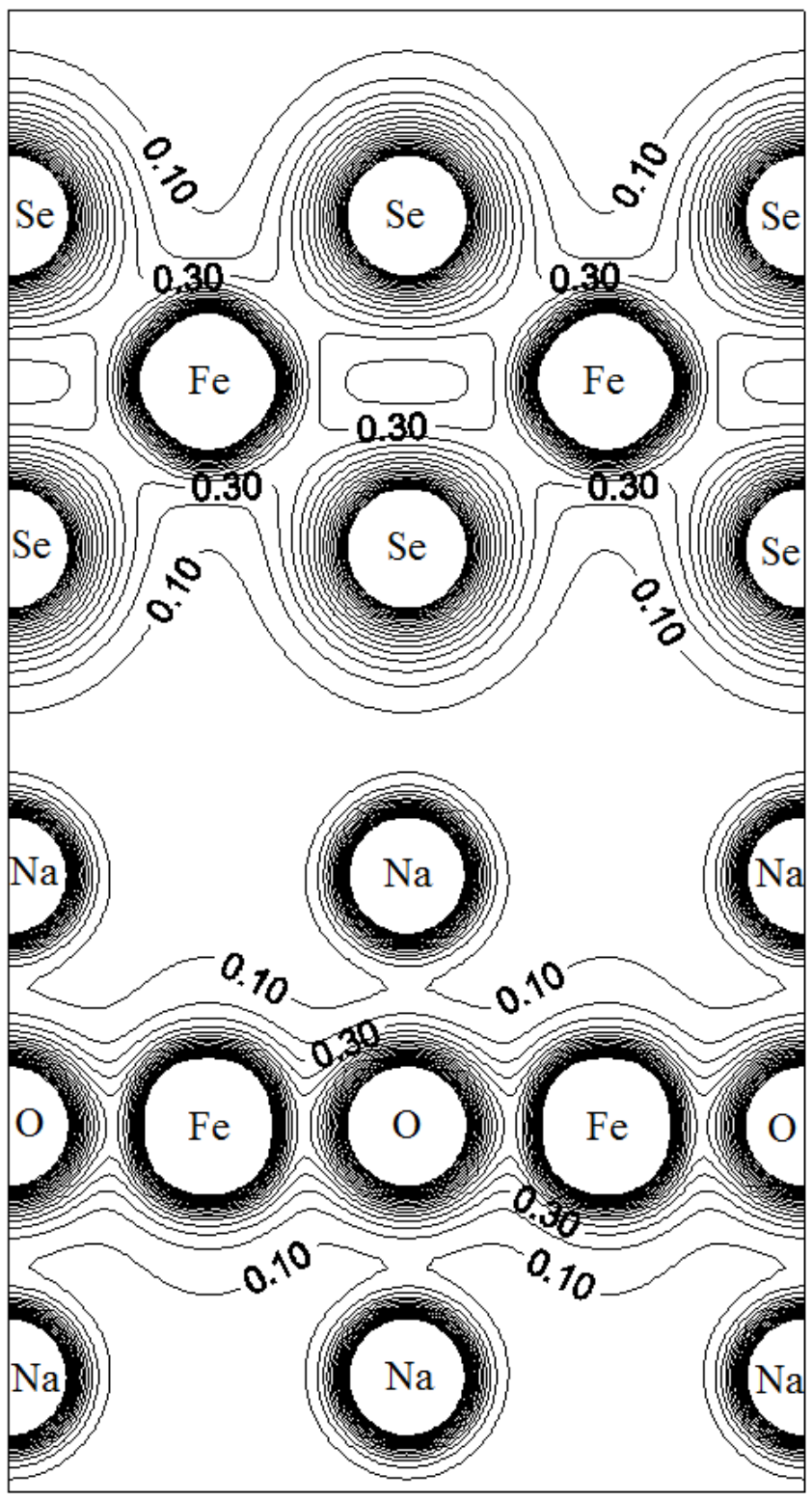

FIG. 2. Valence charge density in (100) plane illustrating directional Fe-Se and Fe-O bonds inside blocks [$Fe_2Se_2O$] of layered $Na_2Ti_2Sb_2O$. The interval between isoelectronic contours is 0.1 e/Å$^3$.

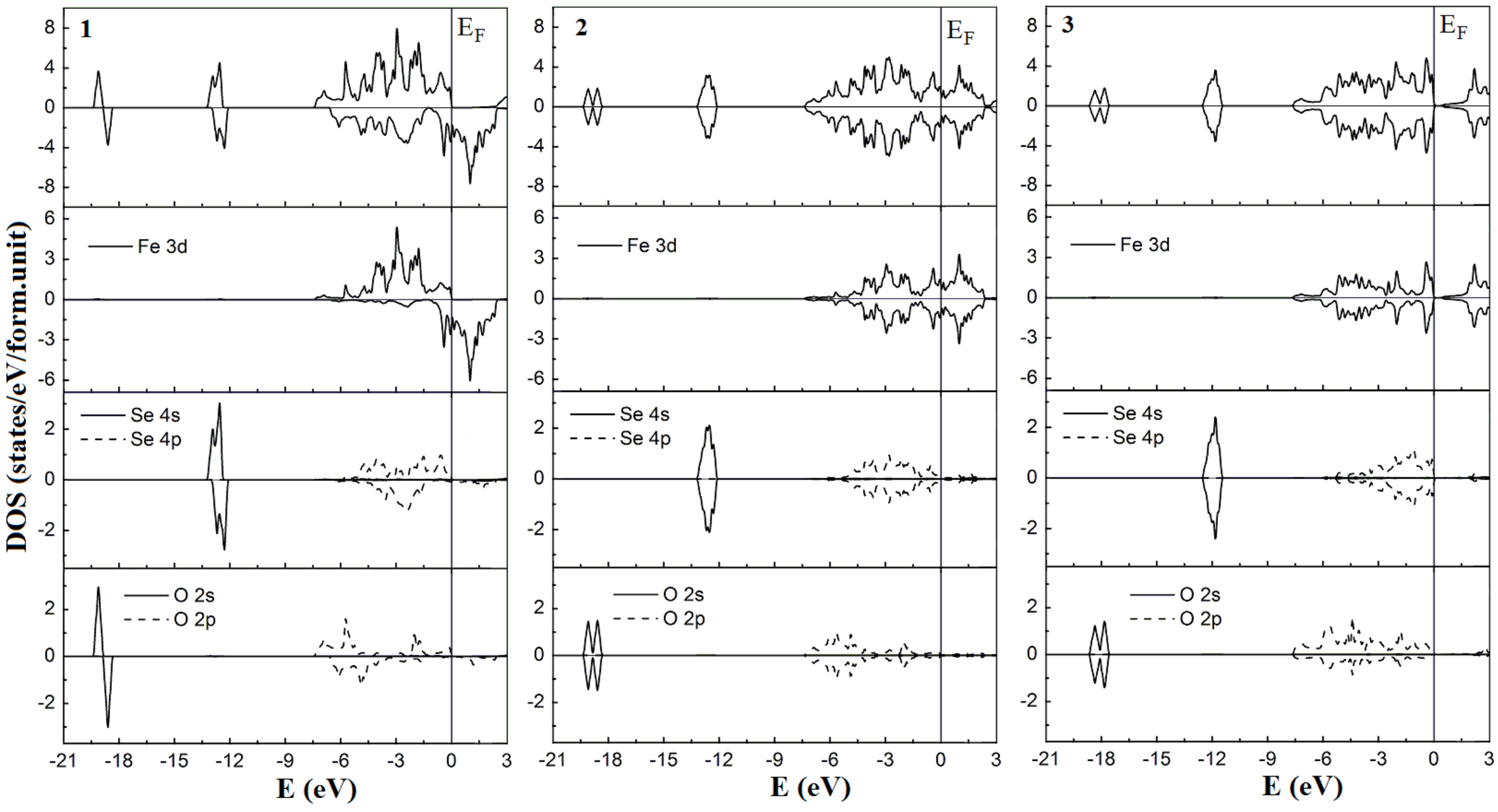

FIG. 3. Spin densities of ferromagnetic (1) and AFM3 (2,3) phases of $Na_2Fe_2Se_2O$ as obtained within GGA (1,2) and GGA+U level (3). The Fermi level is given at 0.0 eV.